\documentclass[journal,final,letterpaper,twoside,twocolumn]{IEEEtran}

\usepackage{amsmath,epsfig}
\usepackage{dsfont}
\usepackage{amsmath}
\usepackage{amssymb}
\usepackage{amsfonts}
\usepackage{graphicx}
\usepackage{amsmath}
\usepackage[noadjust]{cite}
\usepackage[all,poly]{xy}
\usepackage{multirow,balance}
\usepackage{color}      
\usepackage{algorithm,algorithmic,url}

\def\bfa{{\mathbf{a}}}

\def\bfe{{\mathbf{e}}}

\def\bfm{{\mathbf{m}}}
\def\bfn{{\mathbf{n}}}

\def\bfr{{\mathbf{r}}}

\def\bfy{{\mathbf{y}}}

\def\bfA{{\mathbf{A}}}

\def\bfM{{\mathbf{M}}}

\def\bfR{{\mathbf{R}}}

\def\bfY{{\mathbf{Y}}}

\def\bbS{{\mathbb{S}}}

\def\calI{{\mathcal{I}}}

\newcommand{\nopix}{p}
\newcommand{\nbpix}{P}

\newcommand{\nomat}{k}
\newcommand{\nbmat}{K}

\newcommand{\nbband}{L}

\newcommand{\Vpix}[1]{\bfy_{#1}}
\newcommand{\Mpix}{\bfY}

\newcommand{\abond}[2]{a_{#1#2}}
\newcommand{\Vabond}[1]{\bfa_{#1}}
\newcommand{\Mabond}{\bfA}

\newcommand{\Vmat}[1]{\bfm_{#1}}
\newcommand{\Mmat}{\bfM}

\newcommand{\Vres}[1]{\bfr_{#1}}
\newcommand{\Mres}{\bfR}

\newcommand{\Vnoise}[1]{\bfn_{#1}}

\newcommand{\norm}[1]{\left\|#1\right\|}

\newcommand{\Vnrj}{\bfe}

\newcommand{\ve}[1]{ {\mathbf{#1}} }
\def\bal#1{\begin{align}#1\end{align}}
\def\balx#1{\begin{align*}#1\end{align*}}
\def\defequal{\stackrel{\mbox{\footnotesize def}}{=}}
\def\Y{\ve{Y}}
\def\M{\ve{M}}
\def\A{\ve{A}}
\def\R{\ve{R}}
\newcommand{\conv}[1]{\smash{\overset{\scriptscriptstyle\smile}{#1}}}
\newcommand{\conc}[1]{\smash{\overset{\scriptscriptstyle\frown}{#1}}}

\newcommand{\best}[1]{\textcolor[rgb]{0.00,0.00,1.00}{\mathbf{#1}}}
\newcommand{\second}[1]{\textcolor{blue}{#1}}

\title{Nonlinear hyperspectral unmixing with \\robust nonnegative matrix factorization}

\author{C\'edric F\'evotte and Nicolas Dobigeon
\thanks{This work is supported by the ESTOMAT PEPS Project supported by CNRS and by the Hypanema ANR Project n$^\circ$ANR-12-BS03-003.}
\thanks{C. F\'evotte is with Laboratoire Lagrange (CNRS, Universit\'e de Nice Sophia Antipolis and Observatoire de la C\^ote
d'Azur), Parc Valrose, 06108 Nice cedex 2, France. (e-mail:
cedric.fevotte@unice.fr).}
\thanks{N. Dobigeon is with University of Toulouse, IRIT/INP-ENSEEIHT, 2 rue
Camichel, BP 7122, 31071 Toulouse cedex 7, France. (e-mail:
nicolas.dobigeon@enseeiht.fr).} }

\begin{document}
\maketitle
\begin{abstract}
This paper introduces a robust mixing model to describe hyperspectral data resulting from the mixture of several pure spectral signatures. This new model not only generalizes the commonly used linear mixing model, but also allows for possible nonlinear effects to be easily handled, relying on mild assumptions regarding these nonlinearities. The standard nonnegativity and sum-to-one constraints inherent to spectral unmixing are coupled with a group-sparse constraint imposed on the nonlinearity component. This results in a new form of robust nonnegative matrix factorization. The data fidelity term is expressed as a $\beta$-divergence, a continuous family of dissimilarity measures that takes the squared Euclidean distance and the generalized Kullback-Leibler divergence as special cases. The penalized objective is minimized with a block-coordinate descent that involves majorization-minimization updates. Simulation results obtained on synthetic and real data show that the proposed strategy competes with state-of-the-art linear and nonlinear unmixing methods.
\end{abstract}
\section{Introduction}
\label{sec:intro}

Spectral unmixing (SU) is an issue of prime interest when analyzing
hyperspectral data since it provides a comprehensive and meaningful
description of the collected measurements in various application
fields including remote sensing \cite{Asner2002}, planetology
\cite{Themelis2012pss}, food monitoring \cite{Gowen2007} or
spectro-microscopy \cite{Dobigeon2012ultra}. SU consists in
decomposing $\nbpix$ multi-band observations
$\Mpix=\left[\Vpix{1},\ldots,\Vpix{\nbpix}\right]$ into a collection
of $\nbmat$ individual spectra
$\Mmat=\left[\Vmat{1},\ldots,\Vmat{\nbmat}\right]$, called
\emph{endmembers}, and estimating their relative proportions
(or \emph{abundances})
$\Mabond=\left[\Vabond{1},\ldots,\Vabond{\nbpix}\right]$ in each
observation \cite{Keshava2002,Bioucas2012jstars}. Most of the
hyperspectral unmixing algorithms proposed in the signal \& image
processing and geoscience literatures rely on the commonly admitted
linear mixing model (LMM), $\Mpix\approx \Mmat\Mabond$. Indeed, LMM
provides a good approximation of the physical process underlying the
observations and has resulted in interesting results for most
applications. However, for several specific applications, LMM may be
inaccurate and other nonlinear models need to be advocated
\cite{Dobigeon2014}. For instance, in remotely sensed images
composed of vegetation (e.g., trees), interactions of photons with
multiple components of the scene lead to nonlinear effects that can
be taken into account using bilinear models
\cite{Somers2009,Somers2014}. As explained in
\cite{Altmann2011whispers}, several bilinear models have been
proposed \cite{Nascimento2009spie,Fan2009,Halimi2011}, and they
mainly differ by the constraints imposed on the nonlinearity term.
The linear-quadratic model advocated in \cite{Meganem2014} also
incorporates pairwise interactions between the endmembers
components. Conversely, to approximate a large range of second-order
nonlinearities, Altmann \emph{et al.} \cite{Altmann2012ip} introduce
a polynomial post-nonlinear model that has demonstrated its ability
to describe most of the nonlinear effects, in particular in
vegetated areas \cite{Dobigeon2014jstars}. A common feature of these
models is that they all consist in including a supplementary
additive term to the standard LMM, accounting for the
nonlinearities. One major drawback of these models, however, is that
they require to choose a specific form of nonlinearity, and this can
be limiting in practice.\\

In this paper, a new so-called robust LMM (rLMM) is proposed.
Similarly to the nonlinear models detailed above, it is built on the
standard LMM and includes a supplementary additive term that
accounts for nonlinear effects. However, it does not require to
specify an analytical form of the nonlinearity. Instead, nonlinearities are merely treated as
\emph{outliers}. The primary motivation is that the LMM can be
considered as a valid model to describe the majority of the pixels
in a remotely sensed image and, as a consequence, only a
\emph{sparse} number of pixels are affected by nonlinearities. As
such, one of the contributions reported in this article consists in
decomposing the $\nbband\times \nbpix$ matrix of the multi-band
observations as $\ve{Y} \approx \ve{M} \ve{A} + \ve{R}$,
where $\ve{R}$ is a sparse (and nonnegative) residual term
accounting for outliers (i.e., nonlinear effects). To reflect the
assumption that the LMM holds for most of the observed pixels, the sparsity constraint is
imposed at the group-level, i.e., a column of $\ve{R}$ will be
assumed to be either entirely zero or not. The proposed
decomposition relates to \emph{robust nonnegative matrix
factorization} (rNMF) as will be explained in
more details in the sequel of the paper.\\

The article is organized as follows. The rLMM is introduced in more
details in Section \ref{sec:model}. Section \ref{sec:algorithm}
describes a block-coordinate descent algorithm for rLMM estimation.
Experimental results obtained on synthetic data are reported in
Section \ref{sec:simu}. Two real hyperspectral images are
investigated in Section \ref{sec:simu_real}. Section
\ref{sec:conclusion} concludes. This article extends our preliminary conference paper \cite{whispers13} in a significant way. We here generalize the use of the squared Euclidean distance considered in \cite{whispers13} to the more general $\beta$-divergence. Additionally, we show how some of the multiplicative updates obtained heuristically in \cite{whispers13} can be rigorously obtained via majorization-minimization. We also describe a rule of thumb for choosing the value of the penalty weight efficiently. Finally, we provide extended experimental results on synthetical and real data.

\section{Robust linear mixing model} \label{sec:model}

\subsection{Model design}

The proposed rLMM is described by
\begin{equation}
  \label{eq:rlmm}
  \Vpix{\nopix} \approx \sum_{\nomat=1}^\nbmat a_{kp}  \Vmat{\nomat} + \Vres{\nopix} ,
\end{equation}
where $\Vpix{\nopix} = [y_{1p},\ldots,y_{Lp}]^{T} $ denotes the
$\nopix$th pixel spectrum observed in $\nbband$ spectral bands,
$\ve{m}_{k} = [m_{1k},\ldots,m_{Lk}]^{T} $ denotes the $k$th
endmember spectrum,
$\Vabond{\nopix}=\left[a_{1p},\ldots,a_{Kp}\right]^T$ denotes the
abundances associated with the $\nopix$th pixel and $\ve{r}_{p} =
\left[r_{1p},\ldots,r_{Lp} \right]^{T} $ denotes the outlier term
(accounting for nonlinearities). The matrix formulation of Eq. \eqref{eq:rlmm} is given by
\begin{equation}
\label{eq:RLMM}
  \Mpix \approx \Mmat\Mabond + \Mres.
\end{equation}
The approximation symbol in Eqs.~\eqref{eq:rlmm} and \eqref{eq:RLMM} underlies the minimization of a measure of dissimilarity $D(\Mpix|\Mmat\Mabond + \Mres)$, the $\beta$-divergence, that will be introduced in Section~\ref{sec:betadiv}.

The matrices $\ve{Y}$, $\ve{M}$ and $\ve{A}$ are nonnegative by
nature and we take the abundance coefficients to sum to one, i.e.,
\bal{ \label{eq:simplex}
\ve{a}_{p} \in \bbS^{K} \defequal \left\{\boldsymbol{a}\in\mathbb{R}^{\nbmat} \big|\ a_{k}\ge 0,\ \sum_{k=1}^K a_{k} =1 \right\},
}
as commonly assumed in most hyperspectral data
models. In this work, we assume the nonlinear component $\ve{r}_{p}$
to be nonnegative as well, like in the bilinear models of
\cite{Fan2009,Nascimento2009spie,Halimi2011} and the polynomial
model with constructive interferences of \cite{Altmann2012ip}. As
discussed in the introduction, we expect $\ve{r}_{p}$ to be often
zero, i.e., pixels to follow the standard LMM in general. For pixels
where the LMM assumption fails, nonlinearities will become
``active'' and $\ve{r}_{p}$ will become nonzero. This amounts to say
that the energy vector
\begin{equation}
\label{eq:energy}
 \Vnrj =
 \left[\norm{\Vres{1}}_2,\ldots,\norm{\Vres{\nbpix}}_2\right]
\end{equation}
is sparse. In Eq.~\eqref{eq:energy}, $\| \cdot \|_{2}$ denotes the Euclidean norm defined by $\| \ve{x} \|_{2} = \sqrt{\sum_{k}x_{k}^{2}}$. Sparsity can routinely be enforced by $\ell_{1}$-regularization, as done next.

\subsection{Objective function}

In light of previous section, our objective is to solve the minimization problem defined by
\begin{multline} \label{eqn:obj}
\min_{\ve{M},\ve{A},\ve{R}} J(\Mmat,\Mabond,\Mres) = D(\Mpix | \Mmat\Mabond+ \Mres) + \lambda \norm{\Mres}_{2,1} \\
\text{s.t. } \Mmat\ge0,\ \Mabond \ge0,\ \Mres \ge0\ \text{and}\ \norm{\Vabond{\nopix}}_1=1,
\end{multline}
where $\lambda$ is a nonnegative penalty weight, $\ve{A} \ge 0$ denotes nonnegativity of the coefficients of $\ve{A}$, $\| \ve{x} \|_{1} = \sum_{k} |x_{k}|$ and $\| \cdot \|_{2,1}$ is the so-called $\ell_{2,1}$-norm defined by
\begin{equation}
 \norm{\Mres}_{2,1} = \| \ve{e} \|_{1} =  \sum_{\nopix=1}^{\nbpix} \norm{\Vres{\nopix}}_2.
\end{equation}
Eq.~\eqref{eqn:obj} defines a robust NMF problem. Robust NMF is a
nonnegative variant of robust PCA \cite{Candes2009} which has
appeared in different forms in the literature. In
\cite{Sprechmann2012}, the outlier term $\ve{R}$ is nonnegative and
penalized by the $\ell_1$ norm. In \cite{Zhang2011feeeg} and
\cite{Shen2012}, $\ve{R}$ is real-valued and penalized by $\ell_1$
and $\ell_{1,2}$ norms, respectively. In \cite{Kong2011}, the $\ell_{2,1}$ norm of $(\ve{Y} -
\ve{M} \ve{A})$ is minimized (noise free scenario). A so-called robust nonnegative matrix factorization approach was introduced for the reconstruction of reflectance spectra in \cite{BenHamza2006}; however the term ``robust'' there refers to a different feature, namely the use of a data-fitting term (the hypersurface cost function) that is less sensitive to outlier observations than the traditional squared Euclidean distance, for the computation of a regular NMF $\Y \approx \M \A$. Note finally that other articles that have addressed hyperspectral unmixing with regular NMF (i.e., in the standard linear model), e.g., \cite{Pauca2006,Miao2007,Yang2011,Esser2012}.

To the best of our knowledge, the formulation of robust NMF described
by~Eq.~\eqref{eqn:obj}, where $\ve{R}$ is nonnegative and penalized
by the $\ell_{2,1}$ norm (and where the abundances sum to one), is
entirely novel. Furthermore, and most importantly, previous work \cite{Sprechmann2012,Zhang2011feeeg,Shen2012,Kong2011} has only considered
robust NMF with the squared Euclidean distance, i.e., $D(\Mpix |
\Mmat\Mabond+ \Mres) = \| \Mpix - \Mmat\Mabond - \Mres \|_{2}^{2}$
while we give here a more general formulation based on the
$\beta$-divergence, defined next.

\subsection{The $\beta$-divergence} \label{sec:betadiv}

We take the measure of fit in Eq.~\eqref{eqn:obj} to be such that
\bal{
D(\ve{A}|\ve{B}) = \sum_{ij} d(a_{ij} | b_{ij}),
}
where $d(x|y)$ is the $\beta$-divergence between positive scalars $x$ and $y$. The $\beta$-divergence was introduced in various forms in \cite{jor87,basu98,cic10} and has become a standard measure of fit in NMF, see, e.g., \cite{kom07,cic06a,naka10,betanmf}. In this paper we use the following definition:
\begin{equation} \label{eqn:beta}
d_{\beta}(x | y) \defequal
\left\lbrace
\begin{array}{cl}
    \frac{x^{\beta}}{{\beta\,(\beta-1)}} + \frac{y^{\beta}}{\beta} -  \frac{ x\, y^{\beta-1}}{\beta-1}  &  \beta \in \mathbb{R} \backslash \{0,1\}  \\
x\, \log \frac{x}{y} - x + y & \beta = 1 \\
\frac{x}{y} - \log \frac{x}{y} - 1 & \beta = 0
\end{array}
\right. .
\end{equation}
The limiting cases $\beta = 0$ and $\beta = 1$ correspond to the
Itakura-Saito and Kullback-Leibler divergences, respectively. The
squared Euclidean distance is obtained for $\beta=2$, i.e.,
$d_{\beta=2}(x|y)=(x-y)^2/2$. The parameter $\beta$ essentially
governs the assumed statistics of the observation noise and can
either be fixed  or learnt from training data by cross-validation.
Under certain assumptions, the $\beta$-divergence can be mapped to a
log-likelihood function for the Tweedie distribution
\cite{twee84,jor87,ardnmfj}, parametrized with respect to its mean.
In our setting, this translates into $\text{E}[\ve{Y}| \ve{M},
\ve{A}, \ve{R}] = \ve{M} \ve{A} + \ve{R}$. In particular, the values
$\beta =0,1,2$ underlie multiplicative Gamma noise, Poisson noise
and  Gaussian additive noise, respectively. The $\beta$-divergence
offers a continuum of noise statistics that interpolates between
these three specific cases. A noteworthy property of the
$\beta$-divergence is its behavior with respect to scale. Indeed,
let $\lambda \in \mathbb{R}^{+}$, then we have $d_{\beta}(\lambda x|
\lambda y) = \lambda^{\beta} d_{\beta}(x|y)$. This means that the
data-fitting term will rely more heavily on large (resp., small)
coefficients in $\Y$ for $\beta > 0$ (resp., $\beta <0$); see a more
detailed discussion in \cite{neco09}.

\section{Block-coordinate descent algorithm}
\label{sec:algorithm}

In order to solve the rNMF minimization problem defined at Eq.~\eqref{eqn:obj},
we present an iterative block-coordinate descent algorithm that
updates each of the parameters $\M$, $\A$ and $\R$ in turn. Each
parameter is updated conditionally upon the current value of the
other parameters and such that the objective function is decreased.
This is the updating scheme employed by virtually all NMF
algorithms. Unfortunately, given the non-convexity of the objective
function $J(\M,\A,\R)$, this strategy can return local solutions and
proper initialization is required. This will be addressed in Section~\ref{sec:simu}. The updates of the parameters are
described next. In short, the parameters $\M$ and $\R$ are updated
via majorization-minimization (MM). The parameter $\A$ is updated using
a heuristic scheme that has proven to work well in the literature.
All the updates turn out to be ``multiplicative'', i.e., such that
the new update is obtained by term-to-term multiplying the previous
update by a nonnegative matrix, hence automatically preserving the
nonnegativity of the estimates through iterations. The resulting algorithm has linear complexity $\mathcal{O}(LKP)$ (in flops) per iteration.

\subsection{Update of the endmember spectra $\M$} \label{sec:upM}

Updating $\M$ given the current values of $\A$ and $\R$ involves
solving the following minimization problem \bal{ \min_{\ve{M}} \,
C(\M) = D(\Y|\M \A + \R) \ \, \text{s.t. } \ \M \ge 0. } When $\R =
\ve{0}$, this problem boils down to updating the dictionary matrix
in NMF with the $\beta$-divergence. MM algorithms have been designed
for that purpose in \cite{naka10,betanmf}. In this section, we
extend the MM approach to the case where $\R \ge \ve{0}$.  Denote by
$\tilde{\M}$ the estimate of $\M$ at current iteration. The first
step of MM consists in building an upper bound $G(\M| \tilde{\M})$
of $C(\M)$ which is tight for $\M = \tilde{\M}$, i.e., $C(\M) \le
G(\M|\tilde{\M}) $ for all $\M$ and $C(\tilde{\M}) =
G(\tilde{\M}|\tilde{\M}) $. The second step consists in minimizing
the bound with respect to (w.r.t) $\M$, producing a valid descent
algorithm. Indeed, at iteration $i+1$, it holds by construction that
$C(\M^{(i+1)}) \le G(\M^{(i+1)}|\M^{(i)}) \le G(\M^{(i)}|\M^{(i)}) =
C(\M^{(i)}) $. The bound $G({\M}|\tilde{\M})$ will be referred to as
\emph{auxiliary function}.

The auxiliary function obtained in \cite{betanmf} relies on a
convex-concave decomposition of $d(x|y)$ and we follow a similar
approach here. The $\beta$-divergence can always be decomposed as
$d(x|y) = \conv{d}(x|y) + \conc{d}(x|y) + \text{cst} $ where $
\conv{d}(x|y)$ and $\conc{d}(x|y)$ are respectively convex and
concave functions of $y$ and $\text{cst} $ is constant w.r.t $y$. Such
a decomposition is not unique; we give a ``natural'' decomposition
in Table~\ref{tab:beta}. It follows that $C(\M)$ can be decomposed
as the sum of a convex term $\conv{C}(\M)$, a concave term
$\conc{C}(\M)$ and a constant term, such that \bal{
\conv{C}(\M) & = \sum_{lp} \conv{d}(y_{lp} | \sum_{k} m_{lk} a_{kp} + r_{lp} ), \label{eqn:Cconv} \\
\conc{C}(\M) & = \sum_{lp} \conc{d}(y_{lp} | \sum_{k} m_{lk} a_{kp}
+ r_{lp} ). }

\begin{table}[t]
\centering
\renewcommand{\arraystretch}{1.6}
\caption{Differentiable convex-concave decomposition of the
$\beta$-divergence and MM update exponents.}
\begin{tabular}{|c||c|c|c|c|}
\hline
 & $\conv{d}(x|y)$  & $\conc{d}(x|y)$ &  $\gamma(\beta)$ & $\xi(\beta)$ \\
\hline \hline
$\beta < 1 $ and $\beta \not= 0$ &  $- \frac{1}{\beta-1} x \,y^{\beta -1} $ & $\frac{1}{\beta} y^\beta  $ & $\frac{1}{2-\beta} $ & $\frac{1}{3-\beta} $ \\
\hline
$\beta = 0$ & $ x \, y^{-1} $ & $\log{y} $ & $ \frac{1}{2} $ & $\frac{1}{3}$\\
\hline
$1 \le \beta \le 2$ & $d(x|y)$ & 0 & 1 & $\frac{1}{3-\beta} $ \\
\hline
$\beta >2$  &  $\frac{1}{\beta} y^\beta  $ & $- \frac{1}{\beta-1} x \, y^{\beta -1} $ & $\frac{1}{\beta-1} $ & $\frac{1}{\beta-1} $ \\
\hline
\end{tabular}
\label{tab:beta}
\end{table}

From there, $\conv{C}(\M)$ can be majorized using 
Jensen's inequality, as follows. Let us denote $\tilde{y}_{lp} =
\sum_{k} \tilde{m}_{lk} a_{kp} + r_{lp}$ the data approximation
formed with the current iterate $\tilde{\M}$ (and recall that $\A$
and $\R$ are here treated as constants). Then, define for
$k=1,\ldots,K$, $\tilde{\lambda}_{lkp} =
\tilde{m}_{lk}a_{kp}/\tilde{y}_{lp}$ and for $k = K+1$,
$\tilde{\lambda}_{lkp} = r_{lp}/\tilde{y}_{lp}$. By construction, we have
$\sum_{k=1}^{K+1} \tilde{\lambda}_{lkp} =1 $. Then, using definition
of convexity, we have
\bal{
\conv{C}(\M) &= \sum_{lp} \conv{d} \left( y_{lp} | \sum_{k}  \tilde{\lambda}_{lkp} \frac{ m_{lk} a_{kp}}{\tilde{\lambda}_{lkp}} + \tilde{\lambda}_{l(K+1)p} \frac{r_{lp}}{\tilde{\lambda}_{l(K+1)p}} \right) \nonumber \\
& \le \sum_{lp} \left[ \sum_{k=1}^{K}  \tilde{\lambda}_{lkp} \conv{d}\left(y_{lp} |  \frac{m_{lk} a_{kp}}{\tilde{\lambda}_{lkp}} \right) \right. \nonumber \\
&  \quad  \quad \quad \quad \quad \quad \quad \quad \quad + \left. \tilde{\lambda}_{l(K+1)p} \conv{d}\left(y_{lp} | \frac{r_{lp}}{ \tilde{\lambda}_{l(K+1)p}} \right) \right] \nonumber \\
&= \sum_{lp} \left[ \sum_{k=1}^{K} \frac{\tilde{m}_{lk} a_{kp} }{\tilde{y}_{lp}} \conv{d}\left(y_{lp}| \tilde{y}_{lp}\frac{m_{lk}}{\tilde{m}_{lk}}\right) + \frac{r_{lp}}{\tilde{y}_{lp}} \conv{d}(y_{lp}|\tilde{y}_{lp}) \right] \nonumber \\
& \defequal \conv{G}(\M| \tilde{\M})
}
The auxiliary function essentially ``breaks'' the sum over $k$ in Eq.~\eqref{eqn:Cconv} to make the optimization over $\M$ separable w.r.t its entries $m_{lk}$. 

Thanks to its concavity, $\conc{C}(\M)$ can be majorized by a first-order approximation at $\M = \tilde{\M}$ (the tangent inequality), leading to
\begin{multline}
\conc{G}(\M| \tilde{\M}) = \conc{C}(\tilde{\M}) + \sum_{lp}  \conc{d}'(y_{lp} | \tilde{y}_{lp}) \sum_{k} a_{kp} (m_{lk} - \tilde{m}_{lk}),
\end{multline}
where $\conc{d}'(x|y)$ denotes the derivative of $\conc{d}(x|y)$ w.r.t $y$.

An upper bound to $C(\M)$ is finally obtained (up to constant terms) by adding $\conv{G}(\M| \tilde{\M})$ and $\conc{G}(\M| \tilde{\M})$. Skipping details for brevity, the resulting function can be minimized in closed-form w.r.t $\tilde{\M}$, resulting in the following update
\bal{ \label{eqn:upM}
m_{lk} = \tilde{m}_{lk} \left( \frac{\sum_{p} a_{kp} y_{lp} \tilde{y}_{lp}^{\beta-2}  }{\sum_{p} a_{kp} \tilde{y}_{lp}^{{\beta-1}} } \right)^{\gamma(\beta)},
}
where $\gamma(\beta)$ is given in Table~\ref{tab:beta} and we recall that $\tilde{y}_{lp} = \sum_{k} \tilde{m}_{lk} a_{kp} + r_{lp}$ is the data approximation at current iteration. Note that we observed in practice that setting $\gamma(\beta) =1$ for every value of $\beta$ still reduced the objective function at every iteration and produced faster convergence. This may be interpreted as over-relaxation of the MM update, see \cite{betanmf} for further discussion on this subject.

\subsection{Update of the outlier term $\R$} \label{sec:upR}
Updating $\R$ given the current values of $\M$ and $\A$ involves
solving the following minimization problem \bal{ \min_{\R} C(\R) =
D(\Y|\M \A + \R) + \lambda \|\R \|_{2,1} \ \text{s.t. } \ \R\ge0. }
The data-fitting term may be majorized using a convex-concave
decomposition of $D(\Y|\M \A + \R)$ exactly as we did in
Section~\ref{sec:upM}. Denote by $\tilde{\R}$ the current update of
$\R$, $s_{lp} = \sum_{k} m_{lk} a_{kp} = [\M \A]_{lp}$ the low-rank
component and $\tilde{y}_{lp} = s_{lp} + \tilde{r}_{lp}$ the current
data approximation.\footnote{The same notation $\tilde{y}_{lp}$ is
used for $\tilde{y}_{lp} = \sum_{k} \tilde{m}_{lk}a_{kp} + r_{lp}$
in Section~\ref{sec:upM} and for $\tilde{y}_{lp} = \sum_{k}
{m}_{lk}a_{kp} + \tilde{r}_{lp}$  in Section~\ref{sec:upR}. Our
intent is too avoid the use of too many notations and the definition
of $\tilde{y}_{lp}$ should be clear from context (i.e., which
parameter update is considered).} Then, applying the Jensen and
tangent inequalities to the convex and concave parts, respectively,
we obtain \bal{
D(\Y|\M \A + \R) \le \sum_{lp} \left[ \frac{\tilde{r}_{lp}}{\tilde{y}_{lp}} \conv{d}(y_{lp}|\tilde{y}_{lp} \frac{r_{lp}}{\tilde{r}_{lp}}) + \frac{s_{lp}}{\tilde{y}_{lp}} \conv{d}(y_{lp}|\tilde{y}_{lp}) \right] \nonumber\\
 + \sum_{lp} \left[ \conc{d}(y_{lp}| \tilde{y}_{lp}) +  \conc{d}'(y_{lp}| \tilde{y}_{lp})(r_{lp}-\tilde{r}_{lp}) \right]. \label{eqn:FR}
}
Denote by $F(\R|\tilde{\R})$ the right-hand side of Eq.~\eqref{eqn:FR}. An auxiliary function for $C(\R)$ may simply be obtained as $G(\R|\tilde{\R}) = F(\R|\tilde{\R}) + \lambda \|\R \|_{2,1} $. However, this specific auxiliary function is not amenable to optimization w.r.t $\R$ (no closed-form solution). Hence, the first step of our strategy is to majorize the penalty function $\|\R \|_{2,1}$ as well. By concavity of the square-root function, we may write
\bal{ \label{eqn:majR}
\|\R\|_{2,1} \le \frac{1}{2} \sum_{p} \left( \frac{ \|\ve{r}_{p} \|_{2}^{2} }{ \| \tilde{\ve{r}}_{p} \|_{2} } + \| \tilde{\ve{r}}_{p} \|_{2} \right).
}
Equation~\eqref{eqn:majR} essentially replaces $\sqrt{\sum_{l} r_{lp}^{2}}$ by a quadratic tight upper-bound that involves $\sum_{l} r_{lp}^{2}$, with the effect of decoupling the spectral bands from within the square root. Unfortunately, the resulting auxiliary function is not yet amenable to optimization. As such, from here our approach closely follows \cite{ardnmfj}. For $\beta > 2$, we may majorize $r_{lp}^{2}$ by a monomial of degree $\beta$, matching the monomial of highest degree in $F(\R|\tilde{\R})$ (see Table~\ref{tab:beta}). For $\beta \le 2$, the reverse is done: the leading monomial in $F(\R|\tilde{\R})$ is now of degree lower than 2 and may as such be majorized by a quadratic term, matching the quadratic upper bound of the penalty function; see Section 4.1 in \cite{ardnmfj} for more details. This strategy leads to the following update
\bal{ \label{eqn:upR}
r_{lp} = \tilde{r}_{lp} \left( \frac{ y_{lp} \tilde{y}_{lp}^{\beta-2} }{ \tilde{y}_{lp}^{\beta-1} + \lambda \frac{\tilde{r}_{lp}}{\| \tilde{\ve{r}}_{p} \|_{2}} } \right)^{\xi(\beta)},
}
where $\xi(\beta)$ is the exponent given in Table~\ref{tab:beta}. Again, we observed in practice that setting $\xi(\beta) =1$ for every value of $\beta$ still reduced the objective function at every iteration and produced faster convergence.

\subsection{Update of the abundances $\A$}
Updating $\A$ given the current values of $\M$ and $\R$ involves solving the following minimization problem
\bal{ \label{eqn:opA}
\min_{\A} C(\A) = D(\Y|\M \A + \R)  \ \text{s.t. } \; \A \ge0 \; \text{and} \; \forall p, \| \ve{a}_{p} \|_{1} =1 .
}
The sum-to-one constraint on the abundances induces an extra difficulty as compared to the optimization problems involved by the updates of $\M$ and $\R$. In some cases such a constraint can be handled using Lagrange multipliers, but this approach does not succeed in our setting, except in the special case $\beta=1$, corresponding to the generalized Kullback-Leibler divergence. We hence resort to another common approach based on a change of variable. We introduce the variable $\ve{U}$ to be a nonnegative matrix of dimension $K \times P$ and set
\bal{
a_{kp} = \frac{u_{kp}}{ \| \ve{u}_{p} \|_{1}}.
}
The optimization problem of Eq.~\eqref{eqn:opA} is turned into the new optimization problem
\bal{ \label{eqn:opU}
\min_{\ve{U}} C(\ve{U}) & = D\left(\Y \mid \M \left[ \frac{\ve{u}_{1}}{ \| \ve{u}_{1} \|_{1}}, \ldots , \frac{\ve{u}_{P}}{ \| \ve{u}_{P} \|_{1}} \right] + \R \right) \nonumber \\
& \quad \quad \quad \quad \quad \quad \quad \quad \quad \quad \quad \quad \quad   \text{s.t. } \, \ve{U} \ge0
}
which is free from the sum-to-one constraint. This approach has been used for NMF in \cite{egg04}. Unfortunately, we were not able to produce an auxiliary function for the new objective function in~\eqref{eqn:opU} -- in particular because it can no longer be easily decomposed as a convex part and concave part. Instead, we resort to a heuristic commonly used in NMF, see, e.g., \cite{vir07,neco09}, as follows. As it appears, the gradient of $C(\ve{U})$ can be expressed as the difference of two nonnegative functions such that
\bal{
\nabla_{u_{kp}} C(\ve{U}) = \nabla_{u_{kp}}^{+} C(\ve{U}) - \nabla_{u_{kp}}^{-} C(\ve{U}).
}
The heuristic algorithm simply writes
\begin{equation} \label{eqn:upA}
u_{kp} = \tilde{u}_{kp} \frac{  \nabla_{{u}_{kp}}^- C(\tilde{\ve{U}})}{\nabla_{{u}_{kp}}^+ C(\tilde{\ve{U}})}.
\end{equation}
It ensures nonnegativity of the parameter updates provided initialization with a nonnegative value, and produces a descent algorithm in the sense that $u_{kp}$ is updated towards left (resp., right) when the gradient is positive (resp., negative). The algorithm was found experimentally to decrease the value of the objective function at each iteration for every of the many values of $\beta$ that we tried. Denoting $\tilde{s}_{lp} = \sum_{k} m_{lk} \tilde{a}_{kp}$ and $\tilde{y}_{lp} = \tilde{s}_{lp} + r_{lp} $, the update is found to be
\bal{
u_{kp} &= \tilde{u}_{kp} \frac{ \sum_{l} ( m_{lk} y_{lp} \tilde{y}_{lp}^{\beta-2} + \tilde{s}_{lp}\tilde{y}_{lp}^{\beta-1} ) }{ \sum_{l} ( m_{lk} \tilde{y}_{lp}^{\beta-1} + \tilde{s}_{lp}y_{lp}\tilde{y}_{lp}^{\beta-2}) }.
}
The update for $\A$ is then simply $a_{kp} =  {u_{kp}}/{ \| \ve{u}_{p} \|_{1}}$.

As it turns out, the updates~\eqref{eqn:upM},~\eqref{eqn:upR} and~\eqref{eqn:upA} can be implemented in matrix form, as shown in Algorithm~\ref{alg:rnmf}, which recapitulates the overall procedure. In Algorithm~\ref{alg:rnmf}, all operators preceded by a dot `$\cdot$' are entrywise MATLAB-like operations and fraction bars shall be taken term-to-term as well. Additionally, $\ve{1}_{M,N}$ denotes the $M \times N$ matrix with coefficients equal to 1. 

\begin{algorithm}[t]
\begin{algorithmic}
\STATE Initialize $\M$, $\A$ and $\R$
\STATE Set convergence tolerance parameter `tol'
\STATE Set value of $\lambda$
\STATE ${\ve{S}} = \M \A$
\STATE $\hat{\Y} =\ve{S} + \R$
\WHILE{err $\ge$ tol}
\STATE \% Update outlier term $\R$
\STATE
\balx{
\R &\leftarrow {\R} . \left[\frac{\Y. \hat{\Y}^{.(\beta-2)} }{ \hat{\Y}^{.(\beta-1)} + \lambda \, \R \, \text{diag}[ \| \ve{r}_{1}\|_{1}, \ldots, \| \ve{r}_{P}\|_{1}]^{-1}} \right] \\
\hat{\ve{Y}} & \leftarrow \ve{S} + \R
}
\STATE \% Update abundances $\A$
\balx{
\A & \leftarrow \A . \frac{\M^{T} (\Y. \hat{\Y}^{.(\beta-2)}) + \ve{1}_{K,L}(\ve{S} . \hat{\Y}^{.(\beta-1)})   }{\M^{T} (\hat{\Y}^{.(\beta-1)})  + \ve{1}_{K,L} (\ve{S}.\Y. \hat{\Y}^{.(\beta-2)})} \\
\ve{S} & \leftarrow  \M \A \\
\hat{\ve{Y}} & \leftarrow \ve{S} + \ve{R}
}
\STATE \% Update endmembers $\M$
\balx{
\M & \leftarrow \M . \left[ \frac{ (\Y. \hat{\Y}^{.(\beta-2)}) \A^{T}  }{  (\hat{\Y}^{.(\beta-1)}) \A^{T}} \right] \\
\ve{S} & \leftarrow  \M \A \\
\hat{\ve{Y}} & \leftarrow \ve{S} + \ve{R}
}
\STATE Compute the objective function relative decrease `err' (or any other convergence criterion).
\ENDWHILE
\end{algorithmic}
\caption{Group robust NMF}
\label{alg:rnmf}
\end{algorithm}

\subsection{Setting the value of $\lambda$}

The hyperparameter $\lambda$ controls the trade-off between the
data-fitting term $D(\Y|\M \A + \R)$ and the penalty term $\| \R
\|_{2,1}$. Setting the ``right'' value of $\lambda$ is a difficult
task, like in any other so-called variational approach that involves
a regularization term. We describe in this paragraph a rule of thumb
for choosing $\lambda$ in a plausible range of values. Our approach
is based on the method of moments. It consists in interpreting the
objective function~\eqref{eqn:obj} as a joint likelihood and in
matching the empirical mean of the data with its prior expectation
in the statistical model. As mentioned in Section~\ref{sec:betadiv},
the $\beta$-divergence underlies a Tweedie data distribution such
that $\text{E}[\Y| \M \A + \R] = \M \A + \R $. The term $\lambda \|
\R \|_{2,1}$ can be seen a log-prior term. Using some results from
\cite{Lee2010}, the corresponding prior distribution $p(\ve{r}_{p})$
for each column of $\R$ can be obtained as a scale mixture of
conditionally independent half-Normal distributions, with a Gamma
distribution assigned to the scale parameter. In particular, the
expectation of $r_{pl}$ under this prior can be found to be \bal{
\text{E}[r_{lp}] =  \frac{2}{\sqrt{\pi}}
\frac{\Gamma(K/2+1)}{\Gamma(K/2 + 1/2)} \frac{1}{\lambda} \defequal
\frac{C}{\lambda}. } Let us now assume an unspecified independent
prior model for $\M \A$ but such that $\text{E}[ [\M \A]_{lp}] =
\rho$. Denoting by $\hat{\mu} = (LP)^{-1} \sum {y}_{lp}$ the
empirical data expectation, our approach consists in matching
$\hat{\mu}$ with $\text{E}[ [\M \A]_{lp}] + \text{E}[r_{lp}]$,
leading to \bal{ \label{eqn:lambda} \hat{\lambda} =
\frac{C}{\hat{\mu}-\rho}. } We insist that the latter expression
only provides a handy gross estimate of $\lambda$ that comes with no
statistical guarantee. In particular the estimate of $\lambda$ is
extremely dependent on $\rho$, the prior expectation of $[\M \A]_{lp}$. However, because $\rho$ is lower bounded
by $0$, the estimate of $\lambda$ is lower bounded by $\lambda_{0} =
C/\hat{\mu}$, corresponding to a plausible minimum degree of
sparsity. We used $\lambda = \lambda_{0}$ in the evaluations below
and this was found to provide satisfactory results.

\section{Experiments with synthetic data} \label{sec:simu}

In this section we evaluate the relevance of the rLMM proposed in Section
\ref{sec:model} and the accuracy of the corresponding rNMF algorithm
described in Section \ref{sec:algorithm} using synthetic data.

\subsection{Data generation}

Four $64\times 64$-pixel images
composed of $\nbmat=3 \ \textrm{or} \ 6$ pure spectral components
have been generated according to four different linear and nonlinear
models. The endmember spectra have been extracted from the spectral
library provided with the ENVI software \cite{ENVImanual2003}. The
first image, denoted as $\calI_{\mathrm{LMM}}$, is composed of
pixels following the standard LMM (no nonlinear component)
\begin{equation}
\label{eq:simu_LMM}
  \Vpix{\nopix} = \sum_{\nomat=1}^{\nbmat}
  \abond{\nomat}{\nopix}\Vmat{\nomat} + \Vnoise{\nopix},
\end{equation}
with $\ve{a}_{p} \in \bbS^{K}$. The three other images, denoted $\calI_{\mathrm{NM}}$,
$\calI_{\mathrm{FM}}$ and $\calI_{\mathrm{GBM}}$, are generated as follows. Three fourths of the image pixels are generated according to the LMM in \eqref{eq:simu_LMM} and the remaining fourth is generated according to a model that features nonlinear component interactions. More precisely, the latter pixels are generated according to:
\begin{itemize}
  \item the Nascimento model (NM) \cite{Nascimento2009spie}
        \begin{equation*}
        \Vpix{\nopix} = \sum_{\nomat=1}^{\nbmat}
        \abond{\nomat}{\nopix}\Vmat{\nomat} +
        \sum_{i=1}^{\nbmat-1}\sum_{j=i+1}^{\nbmat}
        b_{{i}{\nopix}} \,\Vmat{i}\odot\Vmat{j}  + \Vnoise{\nopix},
        \end{equation*}
        with
        \bal{
        \begin{bmatrix} \ve{a}_{p} \\ \ve{b}_{p} \end{bmatrix} \in \bbS^{2 K -1}
        }
        and $\ve{b}_{p} = [b_{1p},\ldots,b_{(K-1)p}]^{T}$,
 \item the Fan bilinear model (FM) \cite{Fan2009}
        \begin{equation*}
        \Vpix{\nopix} = \sum_{\nomat=1}^{\nbmat}
        \abond{\nomat}{\nopix}\Vmat{\nomat} +
        \sum_{i=1}^{\nbmat-1}\sum_{j=i+1}^{\nbmat}
        \abond{i}{\nopix}\abond{j}{\nopix} \, \Vmat{i}\odot\Vmat{j} + \Vnoise{\nopix},
        \end{equation*}
        with $\ve{a}_{p} \in \bbS^{K}$,
 \item the generalized bilinear model (GBM) \cite{Altmann2012ip}
\begin{equation*}
  \Vpix{\nopix} = \sum_{\nomat=1}^{\nbmat}
  \abond{\nomat}{\nopix}\Vmat{\nomat} +
  \sum_{i=1}^{\nbmat-1}\sum_{j=i+1}^{\nbmat}
\gamma_{ijp} \, \abond{i}{\nopix}\abond{j}{\nopix} \, \Vmat{i}\odot\Vmat{j}
+ \Vnoise{\nopix},
\end{equation*}
with $\ve{a}_{p} \in \bbS^K$ and where the nonlinear
coefficient $\gamma_{ijp}\in (0,1)$ adjust the bilinear interaction
between the $i$th and $j$th endmembers in the $p$th pixel.
\end{itemize}

In the models introduced above, $\Vmat{i}\odot\Vmat{j}$ stands for the termwise (Hadamard) product. \\

In a first experiment, the four images $\calI_{\mathrm{LMM}}$, $\calI_{\mathrm{NM}}$,
$\calI_{\mathrm{FM}}$ and $\calI_{\mathrm{GBM}}$ have been generated by drawing the abundance coefficients $\ve{a}_{p}$ (or $[ \ve{a}_{p}^{T}\ \ve{b}_{p}^{T}]^{T}$ in the case of $\calI_{\mathrm{NM}}$) randomly and uniformly from their admissible set $\bbS^K$ (or $\bbS^{2K-1}$). In a second experiment, we wanted to evaluate the robustness of the algorithms w.r.t the absence of pure
pixels in the images to be unmixed. To do so, we imposed a cutoff to the abundance coefficients that
removes pure pixels from the observations. As such, in this
case the abundances have been uniformly drawn over a truncated version
of the set defined by \eqref{eq:simplex}, namely
\begin{equation}
\label{eq:simplex2}
        \bbS^K_{0.9} = \left\{\boldsymbol{a}\in\mathbb{R}^{\nbmat} \big|\ a_{k}\ge 0,\ \sum_{k=1}^K a_{k} \leq 0.9
\right\}.
\end{equation}
Finally, in the two experiments the interaction coefficients $\gamma_{ijp}$ appearing in the GBM have been uniformly drawn over the set $(0,1)$ and the additive noise $\ve{n}_{p}$ was chosen white Gaussian
with signal-to-noise ratio $\mathrm{SNR}=30\mathrm{dB}$, which is an
admissible value for most of the real imaging spectrometers.

\begin{table*}[t!]
\caption{Estimation performance in term of
$\textrm{aSAM}\left(\Mmat\right)$ ($\times 10^{-3}$) and
$\textrm{GMSE}^2\left(\Mabond\right)$ ($\times 10^{-3}$). Best
scores appear in blue boldface and second best scores appear in
blue. rNMF is initialized by either VCA or Heylen's method, as
stated between brackets. Refer to text for other details.
\label{tab:results_synth}}
\renewcommand{\arraystretch}{1.3}
\begin{center}
\begin{tabular}{|c|c|c|c|c|c|c||c|c|c|c|c|c|c|}
    \cline{4-14} %
    \multicolumn{3}{c|}{} & \multicolumn{4}{c||}{$\textrm{aSAM}\left(\Mmat\right)$}    & \multicolumn{7}{c|}{$\textrm{GMSE}^2\left(\Mabond\right)$} \\%
    \cline{4-14} %
    \multicolumn{3}{c|}{} & \multirow{2}*{VCA}   & \multirow{2}*{Heylen}    &  rNMF  &  rNMF   & VCA      & \multicolumn{4}{c|}{Heylen}    & rNMF   & rNMF    \\
    \cline{9-12} %
    \multicolumn{3}{c|}{} &                          &                          &    (VCA)     &  (Heylen)     &  +FCLS       &  +NM     &  +FM     &  +GBM   &  +PPNM     &  (VCA)    & (Heylen)     \\
    \hline
     \multirow{8}*{\rotatebox{90}{w/o pure pixels}} & \multirow{4}*{\rotatebox{90}{$R=3$}}
 & $\calI_{\textrm{LMM}}$ &  $   9.71    $&$ 33.86   $&$ \second{7.65}    $&$ \best{6.78}   $&$ 0.10    $&$ \second{0.06}    $&$ 4.32    $&$ 0.07    $&$ 0.07    $&$ 0.07    $&$ \best{0.04}    $  \\ 
 & & $\calI_{\textrm{NM}}$ & $  \best{12.74}   $&$ \second{170.50}  $&$ 202.05  $&$ 256.71  $&$ \best{20.78}   $&$ 73.79   $&$ 78.34   $&$ 96.23   $&$ 91.85   $&$ \second{54.68}   $&$ 82.80   $  \\ 
 & & $\calI_{\textrm{FM}}$ & $  \best{7.12}    $&$ 102.26  $&$ \second{7.55}    $&$ 29.61   $&$ \second{0.84}    $&$ 13.54   $&$ 38.17   $&$ 13.28   $&$ 14.83   $&$ \best{0.71}    $&$ 1.57    $  \\ 
 & & $\calI_{\textrm{GBM}}$ & $ 8.26    $&$ 33.10   $&$ \second{5.69}    $&$ \best{5.02}    $&$ 0.25    $&$ 0.23    $&$ 4.01    $&$ \best{0.07}    $&$ \second{0.08}    $&$ 0.20    $&$ 0.20    $  \\ 
     \cline{2-14} \\[-9.0pt]
     \cline{2-14}
     &  \multirow{4}*{\rotatebox{90}{$R=6$}}
 & $\calI_{\textrm{LMM}}$ &  $   72.33   $&$ 86.10  $&$ \best{23.03}   $&$  \second{24.92}  $&$ 2.74    $&$ 2.90    $&$ 26.03   $&$ 2.84    $&$ 2.76    $&$  \best{0.27}    $&$ \second{0.77}    $  \\ 
 & & $\calI_{\textrm{NM}}$ & $  \second{175.97}  $&$ 249.60  $&$ \best{156.81}  $&$ 238.68  $&$ 53.40   $&$ 49.00   $&$ 75.31   $&$ 74.87   $&$ 70.38   $&$ \best{32.12}   $&$ \second{32.46}   $  \\ 
 & & $\calI_{\textrm{FM}}$ & $  \best{93.28}   $&$ 187.55  $&$ \second{144.74}  $&$ 259.01  $&$ \second{13.67}   $&$ 34.71   $&$ 68.13   $&$ 39.01   $&$ 34.51   $&$ \best{10.41}   $&$ 19.86   $  \\ 
 & & $\calI_{\textrm{GBM}}$ & $ \best{79.24}   $&$ 106.52  $&$ \second{83.08}   $&$ 85.49   $&$ \second{4.55}    $&$ 5.73    $&$ 32.21   $&$ 5.50    $&$ \best{4.31}    $&$ 4.95    $&$ 4.57    $  \\ 
     \cline{1-14}\\[-9pt]
     \cline{1-14}
       \multirow{8}*{\rotatebox{90}{with pure pixels}} &  \multirow{4}*{\rotatebox{90}{$R=3$}}
 & $\calI_{\textrm{LMM}}$ &  $   46.45   $&$ 64.89   $&$ \second{12.37}   $&$ \best{11.91}   $&$ 1.89    $&$ 2.05    $&$ 11.85   $&$ 2.07    $&$ 2.06 $&$ \second{0.16}    $&$ \best{0.14}    $  \\ 
 & & $\calI_{\textrm{NM}}$ & $  \best{46.40}   $&$ \second{176.29}  $&$ 189.76  $&$ 254.58  $&$ \best{23.94}   $&$ 66.17   $&$ 67.18   $&$ 91.05   $&$ 92.78   $&$ \second{52.77}   $&$ 68.75   $  \\ 
 & & $\calI_{\textrm{FM}}$ & $  \second{52.77}   $&$ 214.15  $&$ \best{9.26}    $&$ 239.32  $&$ \second{4.03}    $&$ 107.59  $&$ 104.75  $&$ 96.43   $&$ 115.65  $&$ \best{0.79}    $&$ 20.07   $  \\ 
 & & $\calI_{\textrm{GBM}}$ & $ 48.18   $&$ 58.58   $&$ \second{9.49}    $&$ \best{9.31}    $&$ 2.66    $&$ 2.48    $&$ 5.85    $&$ 1.77    $&$ 1.58    $&$ \second{0.30}    $&$ \best{0.29}    $  \\ 
     \cline{2-14} \\[-9pt]
     \cline{2-14}
      &  \multirow{4}*{\rotatebox{90}{$R=6$}}
 & $\calI_{\textrm{LMM}}$   & $   66.55   $&$ 79.19   $&$ \best{15.72}  $&$ \second{17.67}   $&$ 3.95    $&$ 2.34    $&$ 24.35   $&$ 2.07    $&$ 2.11    $&$ \second{0.63}    $&$ \best{0.39}    $  \\ 
 & & $\calI_{\textrm{NM}}$  & $  224.81  $&$ \second{112.82}  $&$ \best{101.07}  $&$ 265.25  $&$ 63.85   $&$ \best{16.88}   $&$ 68.20   $&$ 39.56   $&$ 41.66   $&$ 30.93   $&$ \second{28.94}   $  \\ 
 & & $\calI_{\textrm{FM}}$  & $  \best{98.16}   $&$ 171.10  $&$ \second{145.41}  $&$ 198.64  $&$ 11.21   $&$ 28.55   $&$ 56.15   $&$ 30.30   $&$ 28.64   $&$ \best{9.28}    $&$ \second{10.20}   $  \\ 
 & & $\calI_{\textrm{GBM}}$ & $ \best{75.21}   $&$ 136.20  $&$ \second{79.34}   $&$ 156.75   $&$\second{ 5.36}    $&$ 16.75   $&$ 40.60   $&$ 16.72   $&$ 16.68   $&$ \best{5.06}    $&$ 10.93   $  \\ 
     \hline
\end{tabular}
\end{center}
\vspace{-0.5cm}
\end{table*}

\subsection{Compared methods}

The four images have been unmixed using rNMF and state-of-the-art algorithms specially designed for the considered models. The state-of-the-art algorithms are two-steps; the endmember matrix $\M$ is estimated in a first step, and then the abundance matrix $\A$ is estimated in a second step, given the endmember estimates (in a so-called ``inversion'' step). In contrast, rNMF performs a joint estimation of $\M$ and $\A$ (and $\R$).

We considered vertex component analysis (VCA) \cite{Nascimento2005} coupled with fully constrained least squares (FCLS) \cite{Heinz2001}. VCA and FCLS are standard endmember extraction and inversion methods designed for the LMM. Besides, we considered the nonlinear endmember extraction technique proposed in \cite{Heylen2011jstsp}, denoted as Heylen's algorithm in what follows, coupled with four different inversion methods designed for various nonlinear models, namely the NM, FM, GBM and the very flexible polynomial post-nonlinear mixing model (PPNM) \cite{Altmann2012ip}. NM inversion is also achieved with FCLS since the NM can be interpreted as a linear mixture of an extended set of endmembers \cite{Nascimento2009spie}. FM inversion is achieved with the algorithm detailed in \cite{Fan2009}, which exploits a first-order Taylor series expansion of the nonlinear term. GBM inversion is achieved with the gradient descent algorithm from \cite{Halimi2011igarss}. Finally, PPNM inversion is addressed with the subgradient-based optimization scheme from \cite{Altmann2012ip}.

rNMF is applied with $\beta=2$ (reflecting the Gaussian additive noise used in the data generation and for fair comparison with the other methods that rely on this assumption as well) and $\lambda = \lambda_0$. We considered initializations by either VCA or Heylen's algorithm. Convergence was stopped when the relative difference between two successive values of the objective function fell under $10^{-5}$.

\subsection{Performance measures}
The performance of the unmixing algorithms was evaluated in terms of
endmember estimation accuracy using the average spectral angle
mapper (aSAM)
\begin{equation*}
  \textrm{aSAM}\left(\Mmat\right) = \frac{1}{\nbmat}
\sum_{\nomat=1}^\nbmat
\mathrm{acos}\left(\frac{\langle\Vmat{\nomat},\hat{\bfm}_{\nomat}\rangle}{\left\|\Vmat{\nomat}\right\|
\left\|\hat{\bfm}_{\nomat}\right\|}\right)
\end{equation*}
and abundance estimation accuracy using the global mean square error
(GMSE)
\begin{equation*}
  \textrm{GMSE}^2\left(\Mabond\right) = \frac{1}{\nbmat\nbpix}\sum_{\nopix=1}^\nbpix \left\|\Vabond{\nopix} -
  \hat{\bfa}_{\nopix}\right\|^2.
\end{equation*}

\subsection{Results and discussion}

The performance measures returned by the unmixing methods are reported in Table
\ref{tab:results_synth}. First, the aSAM values show that the
proposed rNMF algorithm competes favorably with the two considered
state-of-the-art endmember extraction algorithms, namely VCA
and Heylen's algorithm. Initialized by these algorithms, it almost always
improves the endmember estimation accuracy, with
or without pure pixels. Similarly, when analyzing the GMSE related
to abundance estimation, these results demonstrate the flexibility
of the rLMM to model observations coming from various scenarios.
More generally, these results demonstrate the ability of the
rLMM-based unmixing technique to mitigate several kinds of nonlinear
effects while preserving good estimation performance when analyzing
only linear mixtures.

\section{Experiments with real data} \label{sec:simu_real}

In this section we apply rNMF to real hyperspectral datasets and discuss the results.

\subsection{Description of the datasets} \label{subsec:data}

We consider two real hyperspectral images that have been chosen because of availability of partial ground truth.
The first image was acquired over Moffett Field, CA, in 1997, by the
the Airborne Visible Infrared Imaging Spectrometer (AVIRIS)
\cite{JPL_AVIRIS}. Water absorption bands have been removed from the
$224$ spectral bands, leading to $L=189$ spectral bands ranging from
$0.4\mu\textrm{m}$ to $2.5\mu\textrm{m}$ with a nominal bandwidth of
$10$nm. The scene of interest, of size of $50 \times 50$ pixels,
consists of a part of lake and a coastal area composed of soil and
vegetation. This dataset has been previously studied in
\cite{Dobigeon2008,Halimi2011} and, thus, the unmixing
results obtained in the current work can be compared to those
reported in these later references. This dataset will be referred to as the ``Moffett'' image in the following.

The second considered dataset was acquired by the
Hyspex hyperspectral scanner over Villelongue, France, in 2010. The
sensed spectral domain consists of $L=160$ spectral bands ranging
from $0.4\mu\textrm{m}$ to $1.0\mu\textrm{m}$. This image consists
of a forested area where $12$ vegetation species have been identified, during the Madonna project \cite{Sheeren2011}. The sub-image of
interest, of size of $50\times50$ pixels, is known to be mainly
composed of oak and chestnut trees, with an additional unknown
non-planted-tree endmember (referred to as Endm. \#3 in what
follows). This dataset will be referred to as the ``Madonna'' image in the following.

\subsection{Selection of $\beta$ via induction}

Most of the literature in hyperspectral unmixing uses the squared
Euclidean distance for the data-fitting term. This choice is often
driven by common practice rather than physical motivations stemming
from the nature of the data. As mentioned in
paragraph~\ref{sec:betadiv}, divergences are often log-likelihoods
in disguise, and as such, choosing a divergence is akin to making a
noise assumption. Thus, when no obvious physical model supports the
choice of a specific divergence, finding the ``right'' measure of
fit can be seen as a model selection problem. When data with a
ground truth is available for a specific task, a model can be
selected based on its performance for this task. Unfortunately, and
to the best of our knowledge, no such public real hyperspectral data
exists for spectral unmixing, {i.e., with perfectly known
endmember spectra and corresponding abundance coefficients}. Another
way of selecting a model can be based on its ability to predict
unseen or missing data. Such an approach does not require a ground
truth. As such, in this paragraph we study how NMF with the
$\beta$-divergence performs on an interpolation task. Pixels are
randomly removed from real hyperspectral images and those pixels are
reconstructed from the low-rank approximation. {The process
is repeated for various values of $\beta$ and an optimal value of
$\beta$ (in terms of predicting performance) is deduced.}

{More precisely, $25$, $50$ or $75\%$ of the pixels have
been randomly and uniformly removed from the `Moffett'' and
``Madonna'' images described in paragraph \ref{subsec:data}. Then,
}we fitted a low rank approximation $\M \A$ to the remaining pixels
by minimizing \bal{ \label{eqn:missobj} \sum_{(l,p) \in \mathcal{O}}
d_\beta(y_{lp}| [\M \A]_{lp}) } with respect to $\M$ and $\A$, where
$\mathcal{O}$ denotes the set of observed entries. {The
outlier term $\R$ is omitted} in this experiment as it cannot be
{inferred} for the missing entries (there is one outlier
entry per missing data entry and the problem is not identifiable).
The objective function~\eqref{eqn:missobj} can be minimized using a minor modification of the MM algorithm described in Section~\ref{sec:algorithm}, similarly to the factorizations with missing data described in \cite{ho08,betanmf}. 

After estimation, the missing pixels $(l,p)$
belonging to the complement of $\mathcal{O}$ are reconstructed as
$[\M \A]_{lp}$ and the aSAMs values between the original (complete)
data $\Y$ and its reconstruction $\hat{\Y}$ are computed. This
process is repeated for values of $\beta$ from $-1$ to $3$ with a
step-size of $0.5$. For every value of $\beta$, $10$ runs are
considered, corresponding to different random initializations and
different sets of missing pixels. The number of endmembers was set to $K=3$.

\begin{figure}
 \begin{center}
   \includegraphics[width=\linewidth]{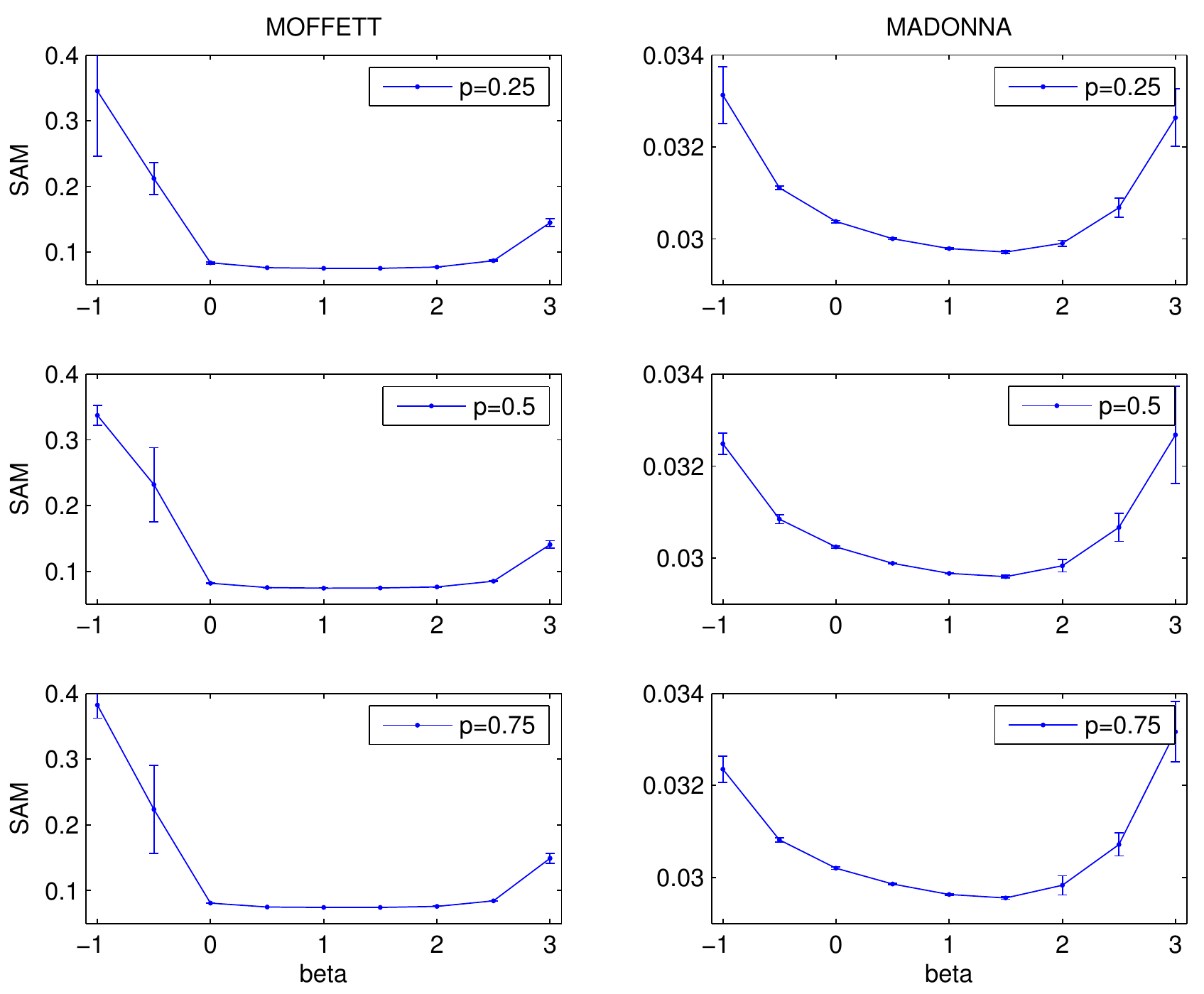}
   \caption{Average SAM values ($\pm$ standard deviation) between the original and reconstructed data over $10$ runs.
   The percentage of observed entries is increased from $25\%$ (top) to $75\%$ (bottom). Left: Moffett image; Right: Madonna image. Best reconstructions are obtained for either $\beta = 1$ or $1.5$.}
   \label{fig:interpol}
   \end{center}
\end{figure}

The results of the experiment are displayed in
Figure~\ref{fig:interpol}. They show that: (1) the choice of $\beta$
matters, (2) best performance is achieved for $\beta = 1$ for the
Moffett image and $\beta = 1.5$ for the Madonna image, with values
of $\beta$ in the $[0,2]$ range yielding sensibly similar
performance. The conclusion of this study is that it can be worth
using alternatives to the standard squared Euclidean distance for
hyperspectral unmixing (such as the KL divergence, corresponding to
$\beta=1$), if it does not come with extra difficulties in the
optimization (as such, the MM algorithm is equally simple to
implement for all values of $\beta$).

\subsection{Robust unmixing results}

\begin{figure}[t]
  \begin{center}
  \begin{tabular}{c}
  \includegraphics[width=0.95\columnwidth]{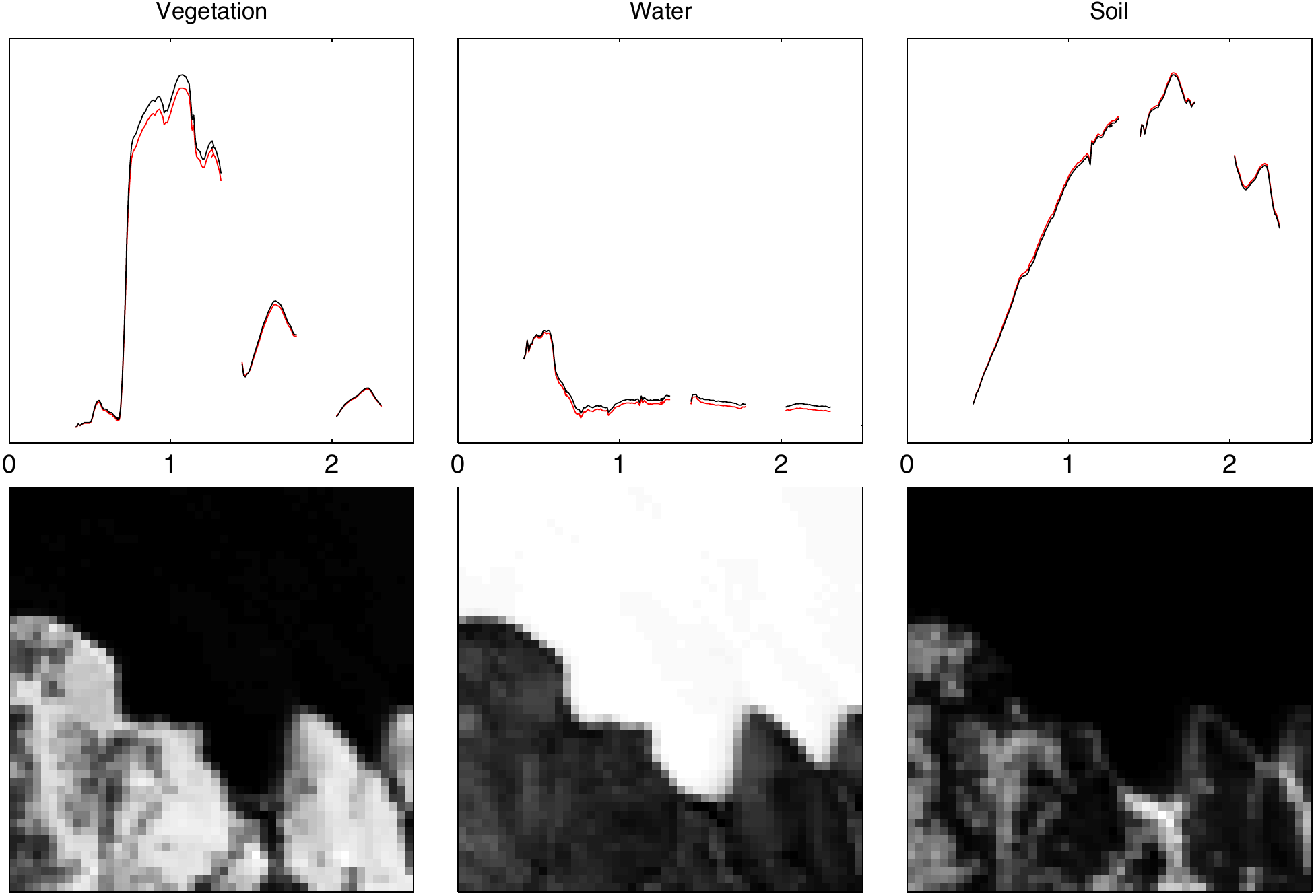} \\
    (a) {\small Moffett } \\
  \\
  \includegraphics[width=0.95\columnwidth]{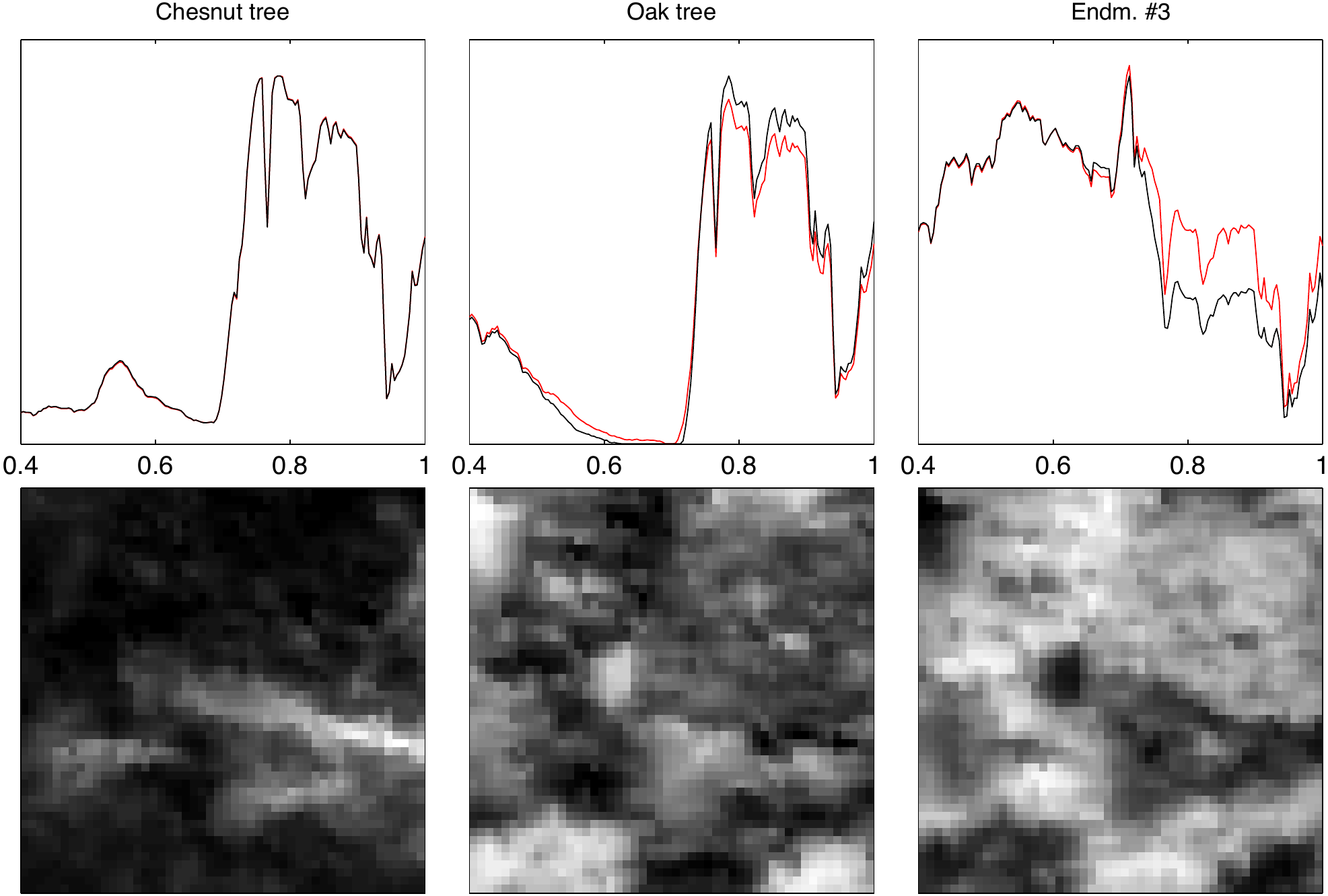} \\
    (b) {\small Madonna} \\
  \end{tabular}
  \end{center}
 \caption{Unmixing results of two real hyperspectral images. Top of each image: endmembers estimated by the proposed rNMF-based unmixing algorithm with $\beta=1$ (red lines) and $\beta=2$ (black lines). Bottom of each image: estimated abundance maps obtained for $\beta=1$; black (resp. white) pixels correspond to absence (resp. presence) of the associated endmembers.
 }
  \label{fig:rnmf}
\end{figure}

In a last experiment, the proposed rLMM-unmixing technique has been applied on the real
Moffett and Madonna images. We used $K=3$ and considered two values of $\beta$, namely $\beta=1$ (Kullback-Leibler divergence) and $\beta=2$ (squared Euclidean distance). The endmember spectra and abundance maps estimated by rNMF are depicted in Fig.~\ref{fig:rnmf}. For conciseness, only the abundance maps obtained with $\beta =1$ are displayed as the results for $\beta =2$ were visually very similar.

\begin{figure}[t]
  \begin{center}
  \begin{tabular}{cc}
  \includegraphics[width=0.45\columnwidth]{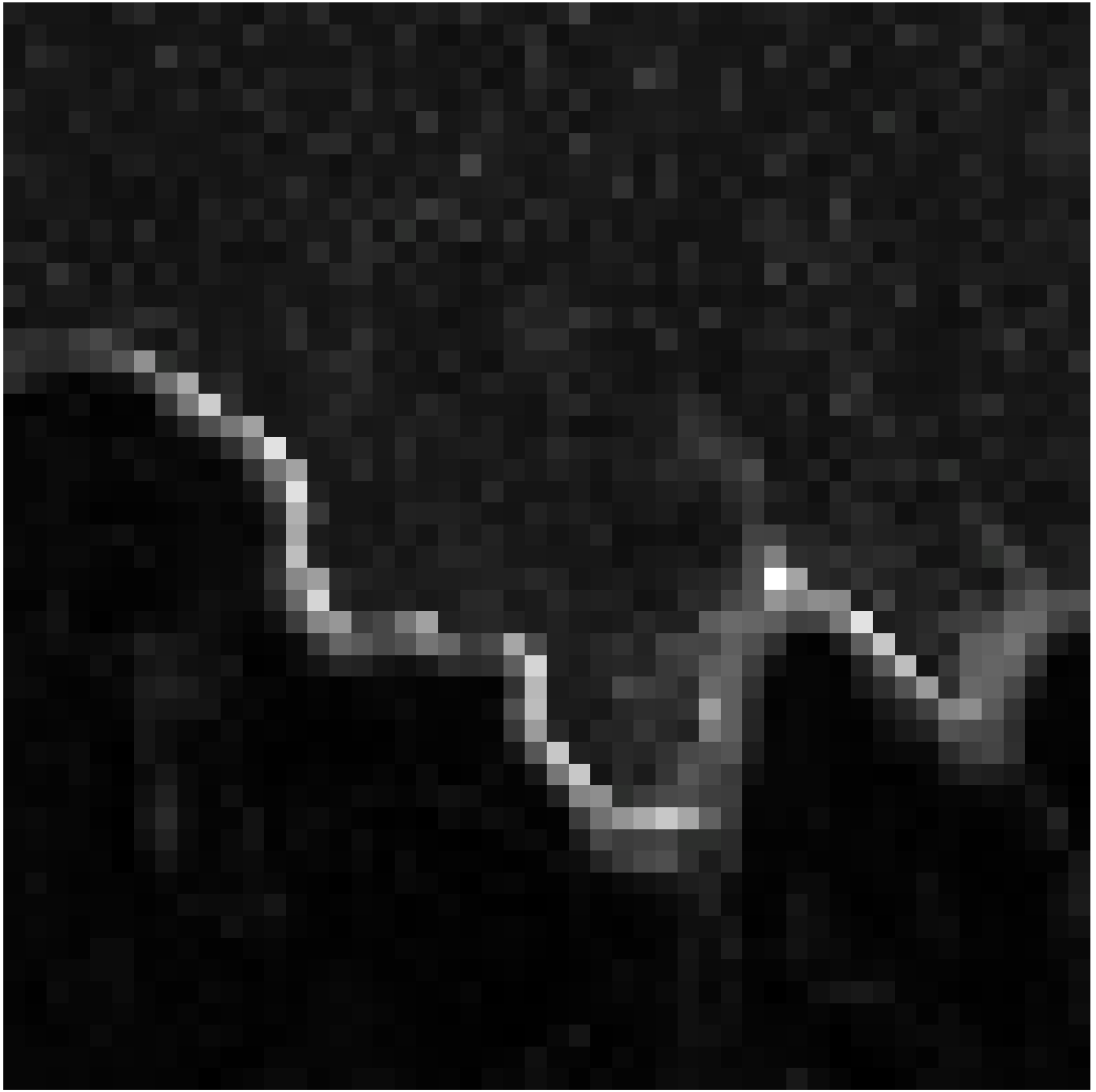} &   \includegraphics[width=0.45\columnwidth]{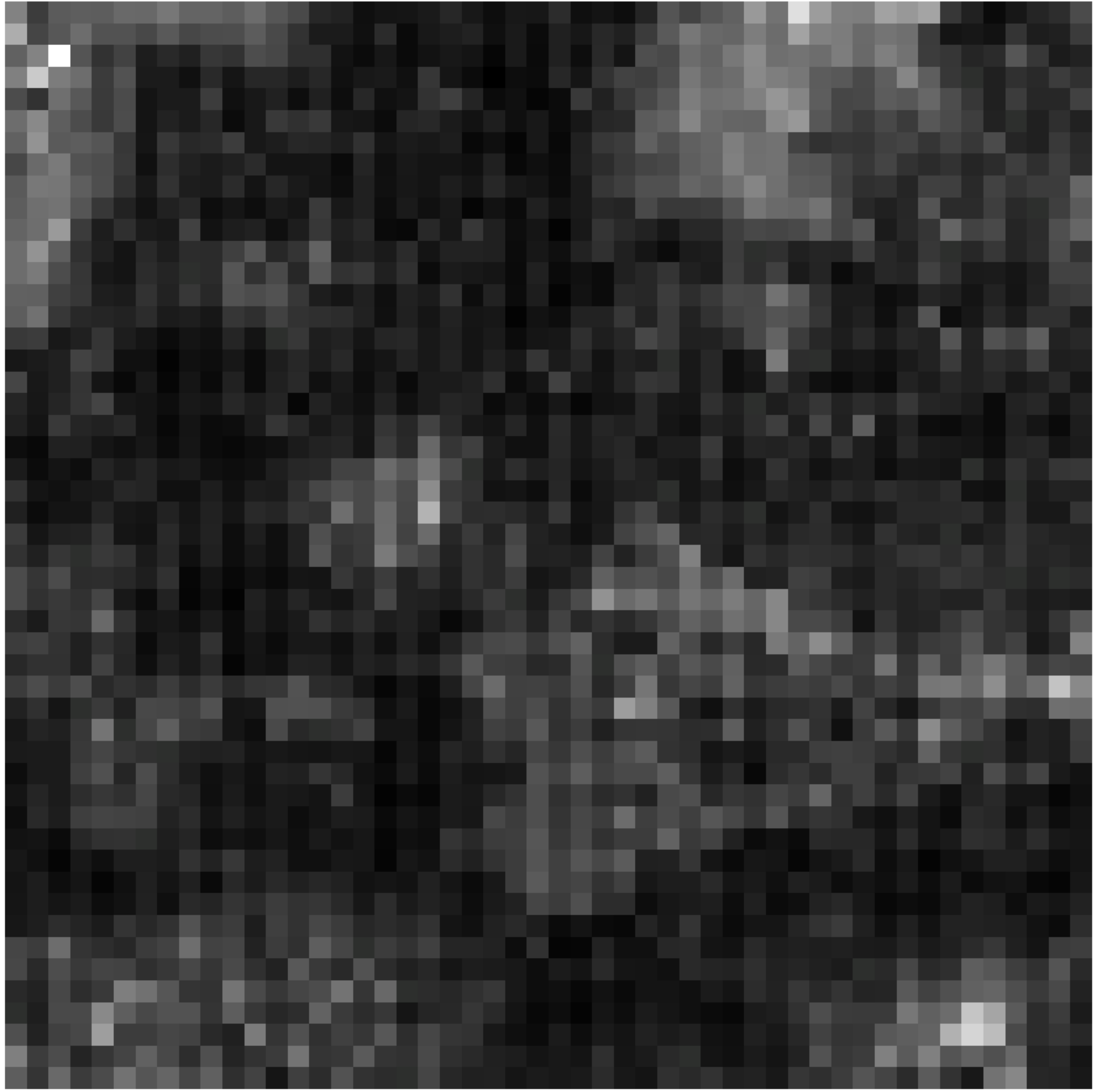} \\
    (a) {\small Moffett } &     (b) {\small Madonna}
  \end{tabular}
  \end{center}
  \caption{Energy of the nonlinear components returned by rNMF with  $\beta=1$. Dark (resp. light) pixels correspond to small (resp. large) values.}
  \label{fig:rnmfe}
\end{figure}

The unmixing results are in good agreement with previous results \cite{Dobigeon2008,Altmann2013sp}. However, in addition to the standard description of the data by linearly mixed endmembers, the proposed model also provides information regarding the pixels that cannot be explained with the
standard LMM. As such, Fig.~\ref{fig:rnmfe} displays the energy $\Vnrj = \left[\norm{\Vres{1}}_2,\ldots,\norm{\Vres{\nbpix}}_{2}\right]$ of the
residual component estimated by rNMF. Regarding the Moffett image, the maps demonstrate that most of the pixels of this scene can be accurately described using the LMM. However, some few pixels, mainly located in the lake shore, appear at outliers. These pixels probably correspond to areas where some interactions between several endmembers occur (e.g., water/vegetation, water/soil). Similar results have been already observed in \cite{Besson2011,Halimi2011}, which confirms the relevance of the proposed method. For the Madonna image, the energy map exhibits residual terms that are mainly located in the area occupied by the oak trees and the unknown $3$rd endmember. Furthermore, the image shows regular vertical patterns that are almost surely due to a sensor defect or miscalibration during the data post-processing.

\section{Conclusion} \label{sec:conclusion}

In this paper we have presented a new mixing model to describe hyperspectral
data. This model, denoted as rLMM, extends the standard LMM by
including a residual term that can capture so-called nonlinear
effects. These nonlinear effects are treated as additive and
sparsely active outliers. In contrast with state-of-the-art literature on nonlinear hyperspectral unmixing, our approach does not require the specification of a particular model of nonlinearity.

The resulting unmixing problem was
formulated as a new form of robust NMF problem, for which we
developed a simple and effective block-coordinate descent algorithm that involves multiplicative updates. We provided an effective rule of thumb for setting the value of the penalty weight, which leaves our algorithm virtually free of parameters (only the number of endmembers needs to be specified). Simulations conducted on synthetic and real data have illustrated the relevance of rLMM, which outperformed many unmixing methods designed for various linear and nonlinear models.

\section*{Acknowledgements}
We thank Vincent Y.~F.~Tan and Zhao Renbo from National University of Singapore for discussions related to robust NMF and feedback about this manuscript.

\bibliographystyle{IEEEtran}
\bibliography{bibfiles/strings_all_ref,bibfiles/biblio_rnmf}

\begin{IEEEbiographynophoto}{C\'edric F\'evotte}
received the state engineering and PhD degrees in control and computer science from the \'Ecole Centrale de Nantes, France, in 2000 and 2003, respectively. During his PhD, he was with the Signal Processing Group at the Institut de Recherche en Communication et Cybern\'etique de Nantes (IRCCyN). From 2003 to 2006, he was a research associate with the Signal Processing Laboratory at the University of Cambridge (Engineering Department). He was then a research engineer with the music editing technology start-up company Mist-Technologies (now Audionamix) in Paris. In 2007, he became a CNRS tenured researcher. He was affiliated with LTCI (CNRS \& T\'el\'ecom ParisTech) from 2007 to
2012. Since 2013, he has been with Laboratoire Lagrangre (CNRS, Observatoire de la C\^ote d'Azur \& Universit\'e de Nice Sophia Antipolis). His research interests generally concern statistical signal processing and machine learning, in particular for inverse problems and source separation. He is a member of the IEEE ``Machine Learning for Signal Processing'' technical committee.
\end{IEEEbiographynophoto}

\begin{IEEEbiographynophoto}{Nicolas Dobigeon} received the state engineering degree in electrical engineering from ENSEEIHT, Toulouse, France, and the M.Sc. degree in signal
processing from the National Polytechnic Institute of Toulouse (INP Toulouse), both in June 2004, as well as the Ph.D. degree and Habilitation \`a Diriger des Recherches in Signal Processing from the INP Toulouse in 2007 and 2012, respectively.
He was a Post-Doctoral Research Associate with the Department of Electrical Engineering and Computer Science, University of Michigan, Ann Arbor, MI, USA, from 2007 to 2008. Since 2008, he has been with the National Polytechnic Institute of Toulouse (INP-ENSEEIHT, University of Toulouse) where he is currently an Associate Professor. He conducts his research within the Signal and Communications Group of the IRIT Laboratory and he is also an affiliated faculty member of the Telecommunications for Space and Aeronautics (TeSA) cooperative laboratory.
His current research interests include statistical signal and image processing, with a particular interest in Bayesian inverse problems with applications to remote sensing, biomedical imaging and genomics.

\end{IEEEbiographynophoto}

\end{document}